\documentstyle[twoside,fleqn,espcrc2,epsf,rotate]{article}
\begin{document}
%%%%%%%%%%%%%%%%

\title{Quenched Chiral Behavior of Hadrons with Overlap Fermions
\thanks{Presented by T.\ Draper at
         Lattice 2001, Berlin, Germany.}
\thanks{This work is supported in part by the U.S. Department of Energy
        under grant numbers DE-FG05-84ER40154 and DE-FG02-95ER40907.
       }
}

\author{%
Shao-Jing Dong
  \address[UK]{Department of Physics and Astronomy, 
               University of Kentucky, 
               Lexington, KY 40506, USA},
Terrence Draper
  \addressmark[UK],
Ivan Horv\'{a}th
  \addressmark[UK],
Frank Lee
\address{Center for Nuclear Studies, 
         Dept.\ of Physics, 
         George Washington Univ.,
         Washington, DC 20052, USA}
\address{Jefferson Lab, 
         12000 Jefferson Avenue, 
         Newport News, VA 23606, USA},
and
Jianbo Zhang
\address{CSSM and Dept.\ of Physics and Math.\ Physics,
         Univ.\ of Adelaide, 
         Adelaide, SA 5005, Australia}
       }

\begin{abstract}
We study the quenched chiral behavior of hadrons with the pseudoscalar mass as
low as $\approx 280\,{\rm MeV}$.  We look for quenched chiral logs in the pion
mass, determine the renormalized quark mass, and observe quenched artifacts in
the $a_0$ and $N^*$ propagators.  The calculation is done on a quenched lattice
of size $20^4$ and $a = 0.148(2)\,{\rm fm}$ using overlap fermions and an
improved gauge action.
\end{abstract}

\maketitle

\section{Simulation Details}

Using a $\beta=7.60$ (tree-level tadpole-improved)
L\"{u}scher--Weisz~\cite{Lus85a,Alf95a} gauge action, we study the chiral
properties of hadrons on a $20^4$ lattice with the overlap
fermion~\cite{Neu98a,Nar95a}.  The massive Dirac
operator~\cite{Ale00a,Cap01a,Her01a} is defined so that the tree-level
renormalization of mass and wavefunction is unity.
    \begin{eqnarray}
	D(m_0) 
	& = & 
	(1 - \frac{m_{0}a}{2\rho}) \rho D(\rho) + m_{0}a \\ \nonumber
	& = &
	(\rho + \frac{m_0a}{2}) + (\rho - \frac{m_0a}{2} ) \gamma_5 \epsilon (H)
    \end{eqnarray}
$\epsilon (H) = H /\sqrt{H^2}$, $H = \gamma_5 D_w$, and $D_w$ is the usual
Wilson fermion operator, except with a negative mass parameter $-\rho =
1/2\kappa -4$ in which $\kappa_c < \kappa < 0.25$; we take $\kappa = 0.19$ in
our calculation which corresponds to $\rho = 1.368$.  See~\cite{Don01b} for
more details of the simulation.

%----------------------------------------------------------

\section{Pion Mass and Chiral Logs}

To search for chiral logs, we fit $m_{\pi}^2 a^2$ to~\cite{Ber92a}
\begin{eqnarray} \label{chi_log}
\lefteqn{m_{\pi}^2 a^2 = } \\ \nonumber
& & A m_{0}a \{1 -\delta [\ln(Am_0 a/\Lambda_{\chi}^2 a^2) +1]\} 
+ B m_0^2 a^2
\end{eqnarray}

The best fits which give stable values of A and B and with errors less than
half of the fitted values of $\delta$ for a range of $\Lambda_{\chi} = 0.6$ GeV
to 1.4 GeV are listed in the table.

\begin{table}[hbt] \label{tab:pion_fit}
\vspace{-0.7cm}
\begin{center}
\caption{Quenched chiral log parameter $\delta$ and $\chi^2/NDF$ as fitted
from $m_{\pi}^2 $ in Eq.~(\ref{chi_log}).}
\begin{tabular}{lllll}
\hline
$\Lambda_{\chi}$    & A  & B & $\delta$ & $\chi^2/NDF$  \\
\hline
  0.6   &  1.72(7)  & 3.0(8) & 0.23(7)  & 0.18   \\
  0.8   &  1.42(11) & 3.0(8) & 0.28(11) & 0.18     \\
  1.0   &  1.17(24) & 3.0(8) & 0.34(17) & 0.18    \\
\hline
%\hline
\end{tabular}
\end{center}
\vspace{-0.7cm}
\end{table}

We plot the fit with $\Lambda_{\chi} = 0.8\, {\rm GeV}$ as a solid line in
Fig.~1.  To check if there is indeed a quenched chiral log, we fit $m_{\pi}^2
a^2$ alternatively without the chiral log term and find
%
%\begin{eqnarray}
%m_{\pi}^2 a^2 &=& 1.65(5)\, m_0a + 0.54(21)\, m_0^2 a^2 
%\end{eqnarray}
\begin{eqnarray}
\lefteqn{m_{\pi}^2 a^2 = } \\ \nonumber
& & 1.88(10)\, m_0a - 2.3(11)\, m_0^2 a^2 + 7.8(30)\, m_0^3 a^3
\end{eqnarray}
with $\chi^2/NDF = 0.46$.  A quadratic fit (shown in Fig.~1 as the dotted line)
has $\chi^2/NDF = 0.90$.  Thus the presence of a chiral log term is favored.

\begin{figure}[t] \label{fig:pion_mass}
  \begin{center}
    \epsfxsize=1.5\hsize
    \epsfbox[90 200 730 662]{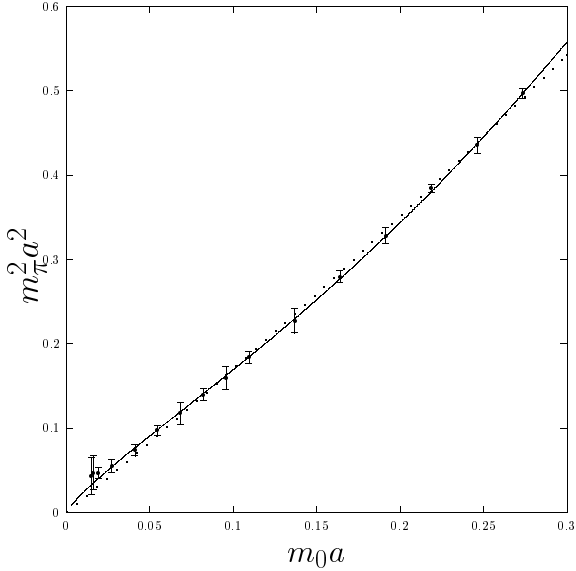}
    \vspace{-1.5cm}
    \caption{$m_{\pi}^2 a^2$ versus $m_0a$ calculated from $G_{A_4P}(\vec{p} =
             0, t)$ with linear plus quadratic fit (dotted line) and chiral log
             fit (solid line) from Eq.~(\ref{chi_log}) with $\Lambda_{\chi} =
             0.8\, {\rm GeV}$.}
    \vspace{-1cm}
  \end{center}
\end{figure}

%----------------------------------------------------------

\section{Quark Masses}

To renormalize our estimates of the quark mass, we follow~\cite{Her01a} to
determine the renormalization constants $Z_M^{-1}=Z_S=Z_P$ and thus $m^{RGI} =
Z_{M}(g_0)m_{0}(g_{0})$: calculate
\begin{eqnarray}
	Z_{M}(g_0) & = & U_M\cdot\frac{1}{(r_{0}m_{0})}|_{(r_{0}m_{\pi})^2=x_{\rm
	ref}}
\end{eqnarray}
using their tabulated values
\begin{eqnarray}
U_{M} & = & \left\{ \begin{array}{ll}
		0.181(6), & x_{\rm ref}=1.5736 \\
		0.349(9), & x_{\rm ref}=3.0 \\
		0.580(12), & x_{\rm ref}=5.0
		\end{array}
		\right.
\end{eqnarray}
where $U_{M}$ is a universal factor which they estimated from published results
of another regularization, namely, ${\cal O}(a)$-improved Wilson.

The value of the bare quark mass $m_{0}a$ that reproduces a given pseudoscalar
mass (through $x_{\rm ref}=(r_{0}m_{\pi})^2$, with $r_{0}/a=4.05(2)$ for our
lattice) is obtained by interpolation of our data for the pion mass.  From this
we obtain $Z_{M}^{-1}=Z_{S}$ at three values of quark mass (the lowest two
quark masses are roughly half-strange and strange) as shown in Fig.~2.

\begin{figure}[t] 
  \begin{center}
    \epsfxsize=1.5\hsize
    \epsfbox[90 200 730 662]{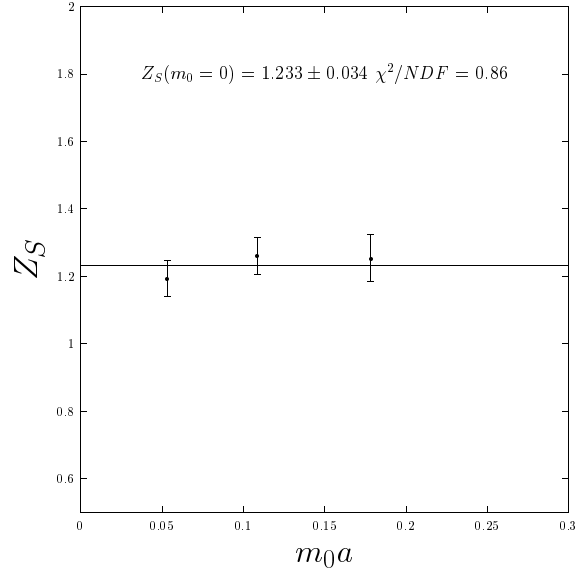}
    \vspace{-1.5cm}
    \caption{Renormalization constant $Z_{M}^{-1}=Z_S$.}
    \vspace{-1cm}
  \end{center}
\end{figure}

The operators and action are ${\cal O}(a)$ improved.  For the axial vector
renormalization constant, $Z_A$, we find small $(m_{0}a)^2$ dependence as
well~\cite{Don01b,Don01a}.  Thus we fit $Z_{M}^{-1}=Z_{S}$ to a constant.
\begin{equation}
	Z_{M}^{-1} = Z_{S}= 1.233(34) \quad\chi^2/NDF=0.86
\end{equation}

From the $(m_{\pi}a)^2$ versus $m_{0}a$ figure, we obtain $m_{0}a$ at the
physical pion mass.  Using $m^{RGI}= m_{0}/Z_{S}$ and, from the 4-loop
calculation with $N_f=0$~\cite{ALP00a}, $Z_{S}^{\overline{MS}}=Z_{S}/0.72076$,
we obtain (very preliminary results) for $(m_{u}+m_{d})/2$.
\begin{eqnarray}
	m^{RGI} 
	& = & 
	7.3(7)(10)\,{\rm MeV} \\
	m^{\overline{MS}}(\mu=2\,{\rm GeV}) 
	& = &
	5.3(6)(7)\,{\rm MeV}
\end{eqnarray}

%----------------------------------------------------------

\section{Chiral Loops in Quenched Propagators}

Bardeen et al.~\cite{Bar01a} present evidence of a strong ``quenched chiral
loop'' (QCL) effect, attributable to an $\eta'$--$\pi$ intermediate state, in
the valence propagation of the scalar, isovector meson ($a_0$).  The effect is
manifested in a negative-metric contribution to the two-point correlation
function which increases for decreasing quark mass.  They were able to uncover
the effect only after using their pole-shifting procedure, which mitigates the
poor chiral properties of the Wilson fermions.  We see this effect clearly with
overlap fermions without the need for any such procedure.

\begin{figure}[t] \label{fig:a0}
  \begin{center}
    \epsfysize=\hsize\leavevmode
    \rotate[l]{\epsfbox{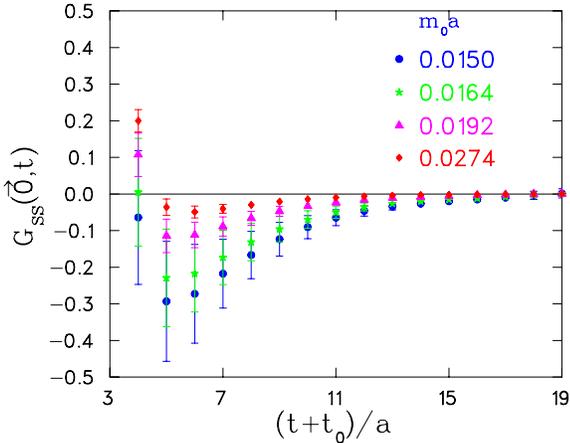}}
    \vspace{-1.5cm}
    \caption{2pt corr.\ func.\ for $a_0$ with bare quark masses
            $m_{0}a=0.0150$, $0.0164$, $0.0192$, and $0.0274$.}
\vspace{-1cm}
  \end{center}
\end{figure}

Fig.~3 shows the two-point scalar-isovector local-local correlators for quark
masses $m_{0}a=0.0150$, $0.0164$, $0.0192$, and $0.0274$; they are plotted on a
linear scale to show the correlator becoming negative at early times
($t/a\approx 1-2$, the source is at $t_0=3$) and then asymptotically
approaching zero from below.

\begin{figure}[t] \label{fig:nucstar}
  \begin{center}
    \epsfysize=\hsize\leavevmode
    \rotate[l]{\epsfbox{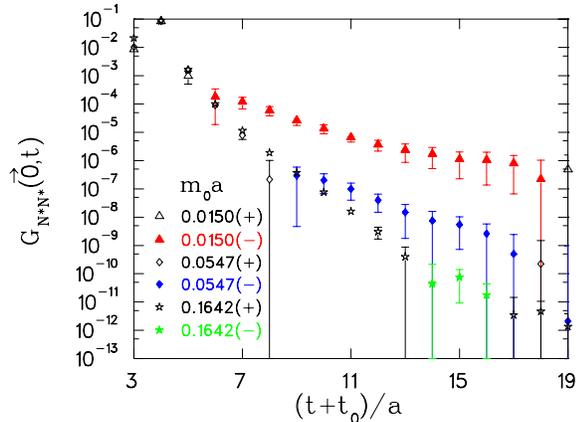}}
    \vspace{-1.5cm}
    \caption{Two-point correlation function of $N^*$ for bare quark masses
             $m_{0}a=0.0150$, $0.0547$, and $0.1642$.  Open (filled) symbols
             and bare masses labeled with a ``($+$)'' (``($-$)'') indicate
             positive (negative) values for the correlation function.}
             \vspace{-1cm}
  \end{center}
\end{figure}

We also see the effect for $N^*(\frac{1}{2}^{-})$.  Fig.~4 shows the two-point
$N^*$ local-local correlator for quark masses $m_{0}a=0.0150$, $0.0547$, and
$0.1642$.  The correlation function becomes negative at successively earlier
times for decreasing mass.  The effect becomes very pronounced at the lighter
masses.  See~\cite{Don01c} for more details.

\section{Summary}

We found evidence for chiral logs in the dependence of the pion mass on quark
mass. (See~\cite{Don01b} for more details.)  We also presented, for the first
time, our very preliminary results for the renormalized light quark mass and
the effect of chiral loops on the $a_0$ and $N^*$ propagators.  In a companion
proceeding~\cite{Don01a}, we study zero modes and their effect on the pion
propagator, and show our results for the axial renormalization constant $Z_A$
and pion axial decay constant $f_{\pi}$, and the presence of chiral logs in the
pseudoscalar decay constant, $f_{P}$.  See~\cite{Don01b} for complete details.

%----------------------------------------------------------

\vfill

\end{document}